\documentclass[12pt]{article}

\usepackage{sbc-template}
\usepackage{graphicx,url}
\usepackage[utf8]{inputenc}  
\usepackage{comment}
\usepackage{amsmath, amssymb, amsfonts}
\usepackage{algorithm}
\usepackage{algorithmic}
\usepackage{listings}
\usepackage{url}
\usepackage{tcolorbox}
\usepackage{cite}
\usepackage{xcolor, soul}
\sethlcolor{green}
\usepackage{ulem}
\usepackage[flushleft]{threeparttable}
\usepackage{listings}
\usepackage{amsmath}
\usepackage{pgfplots}
\pgfplotsset{compat=1.16}
\usepackage{tikz}
\usepackage[hidelinks]{hyperref}
\title{Trust, but Verify: Evaluating Developer Behavior in Mitigating Security Vulnerabilities in Open-Source Software Projects}

\author{Janislley Oliveira de Sousa\inst{1}\inst{,}\inst{2}, Bruno Carvalho de Farias\inst{3}, \\Eddie Batista de Lima Filho\inst{2}\inst{,}\inst{4}, Lucas Carvalho Cordeiro\inst{2}\inst{,}\inst{3}}

\address{Sidia Institute of Science and Technology, Manaus, Brazil
\nextinstitute
    Federal University of Amazonas (UFAM), Manaus, Brazil
\nextinstitute
    University of Manchester, Manchester, United Kingdom
\nextinstitute
    TPV Technology, Manaus, Brazil
  \email{janislley.sousa@sidia.com, bruno.farias@manchester.ac.uk,}
  \email{eddie.filho@tpv-tech.com, lucascordeiro@ufam.edu.br}
}

\begin{document} 

\maketitle

\begin{abstract}
This study investigates vulnerabilities in dependencies of sampled open-source software (OSS) projects, the relationship between these and overall project security, and how developers' behaviors and practices influence their mitigation. Through analysis of OSS projects, we have identified common issues in outdated or unmaintained dependencies, including pointer dereferences and array bounds violations, that pose significant security risks. We have also examined developer responses to formal verifier reports, noting a tendency to dismiss potential issues as false positives, which can lead to overlooked vulnerabilities. Our results suggest that reducing the number of direct dependencies and prioritizing well-established libraries with strong security records are effective strategies for enhancing the software security landscape. Notably, four vulnerabilities were fixed as a result of this study, demonstrating the effectiveness of our mitigation strategies.
\end{abstract}

\section{Introduction}

As widely known, modern software development often employs extensive third-party code from external libraries to save time, which usually comes from open-source software projects and offers numerous advantages, such as transparency, flexibility, and cost-effectiveness. Further analysis also reveals that developers frequently rely on these libraries even when carrying out simple tasks instead of writing their code~\cite{tang2022towards}, and their centralized repositories make download and integration tasks easier, boosting productivity. However, they also present significant risks due to potential problems that can directly impact users~\cite{plate2015impact}.

Indeed, open-source third-party libraries may contain security vulnerabilities~\cite{kula2018developers, pashchenko2020qualitative}. While developers usually review their code for bugs and security issues using specialized tools, e.g., CPPCheck~\cite{marjamaki2013cppcheck}, IKOS~\cite{brat2014ikos}, and ESBMC v7.4~\cite{menezes2024esbmc}, they often skip checking third-party libraries due to the extra effort involved in their evaluation~\cite{kula2018developers}. Going deeper, since a software project may depend on several open-source libraries, which may, in turn, depend on many other libraries in a complex package dependency network, analysis of a software project’s entire dependency tree can become very complex. 

In addition, the C programming language, which is widely used to develop critical open-source projects (e.g., operating systems, device drivers, and encryption libraries), lacks protection mechanisms such as bound checking and memory safety~\cite{lipp2022empirical}, leaving developers responsible for memory and resource management~\cite{berger2019impact}. This way, any lapse may result in undefined behavior, exposing a program to security vulnerabilities. Consequently, developers must be aware of these risks and ensure that pointers in subtraction, addition, and comparison belong to the same memory segment to prevent adverse outcomes. 

Toward this need, the National Cybersecurity Federally Funded Research and Development Center (NCF), operated by the MITRE Corporation, oversees the common vulnerabilities and exposures (CVE) system~\cite{MITRE}, which regularly publishes newly identified open-source vulnerabilities. These vulnerabilities are documented in a comprehensive database with over $237,725$ entries, spanning across different languages, project types, and technologies.

Additionally, although security vulnerabilities, in an isolated manner, already constitute a significant challenge when creating applications with open-source code, they may also be boosted by other factors. Xiao {\it et al.}~\cite{xiao2014social} investigated several social factors impacting developers' adoption decisions based on a multidisciplinary field of study called diffusion of innovations~\cite{wermke2023security}. Their results indicate that security tools can compel developers to build more secure software by detecting and resolving vulnerabilities during the implementation and code review phases. However, it is essential to emphasize the importance of integrating security considerations throughout the entire software development lifecycle to ensure comprehensive protection. Moreover, conditions such as concerning behavior and lack of understanding regarding the consequences of security failures were identified in those whose primary activity is code writing~\cite{assal2018security}. furthermore, while most open-source software projects have large communities contributing to their growth, some are not regularly maintained, which favors security issues~\cite{wermke2022committed}.

Regarding third-party libraries implemented in C language during open-source projects, it is crucial to adopt a critical perspective: developers should thoroughly examine open-source software to identify any vulnerabilities and potential backdoors~\cite{zou2019research}. Even when a project does not use specific vulnerable components directly, an element bundled in some linked package (e.g., third-party library or module) may cause problems and affect others by cascading effects defined as transitive dependency. In other words, examining source code and its documentation is essential to finding software vulnerabilities~\cite{almarimi2020learning}. However, although many developers are mindful of secure-code best practices, there is no guarantee that they will follow all guidelines during development phases or integrate them into software processes. Moreover, some problems may still exist in the available code as it is challenging to detect security risks before software deployment~\cite{gueye2021decade}.

As initial and general perceptions, integrating third-party libraries in open-source projects has become a standard practice to expedite development and leverage existing solutions~\cite{massacci2021technical}. However, this approach introduces significant security challenges~\cite{tang2022towards}. On the one hand, while static analyzers often produce false positives, they can also identify genuine issues that will likely be overlooked during manual code reviews. For instance, in this scenario, Fortify source code analyzer (SCA) \cite{Fortify} excels in detecting vulnerabilities like buffer overflows and structured query language (SQL) injections, while Coverity \cite{Coverity}, a static application security testing (SAST) tool, can identify critical bugs such as memory leaks and concurrency issues. Although both tools may flag non-critical issues, their thorough analysis helps catch significant security flaws that manual review procedures might miss, improving software security and reliability. On the other hand, formal verifiers, supported by mathematical proofs, produce significantly fewer false positives compared to static analysis tools \cite{vsvejda2020interpretation}. This ensures a higher level of accuracy in identifying vulnerabilities. Therefore, developers should balance dismissing all analyzer reports and addressing every single one. 

Ultimately, the primary goal of software quality phases is to ensure software integrity and protect project results from potential threats, regardless of their origin, which includes avoiding risks related to bad practices and common assumptions. We examined various aspects related to the characteristics of the discovered vulnerabilities in the sampled projects' open-source dependencies. Our analysis revealed that developers' behaviors and practices significantly influence the mitigation of security vulnerabilities in third-party libraries within open-source software (OSS) projects. Consequently, this study aims to answer the following research questions:

\begin{itemize}
    \item \textbf{RQ1:} \textit{What are the Common Types and Prevalence of Dependency Vulnerabilities in Open-Source Software Projects?}
    \item \textbf{RQ2:} \textit{How do developers’ behaviors and practices influence the mitigation of security vulnerabilities?}
    \item \textbf{RQ3:} \textit{What is the most effective strategy for mitigating risks from dependency vulnerabilities in open-source software projects?}
\end{itemize}

The remainder of this article is organized as follows: Section~\ref{cap:Background} describes the key concepts used in this study, including the tools and techniques employed for vulnerability detection and analysis. Next, Section~\ref{sec:Methodology} shows the methodology defined to execute the experiments. Section~\ref{cap:results} provides a detailed analysis of the identified vulnerabilities in various OSS projects, discusses how these vulnerabilities are managed by developers, and presents the outcomes of the remediation efforts, including the specific fixes applied. Lastly, Section~\ref{cap:conclusion} summarizes our findings, discusses the implications of developer behaviors on security practices, and offers recommendations for mitigating security vulnerabilities in OSS projects.

\section{Background}
\label{cap:Background}

Software developers frequently use open-source libraries to speed up development cycles, but these libraries can contain security vulnerabilities, leading to high-profile incidents. Besides, as the use of open-source libraries grows, managing and mitigating these dependency vulnerabilities becomes increasingly important~\cite{prana2021out}. In that sense, testing is inevitable. However, it is important to understand that software quality protocols are not simple evaluation sessions. Indeed, the complete process for software verification usually includes vulnerability identification, confirmation, code analysis, and code repair (e.g., patch application and merge requests), which may even be extended (e.g., robustness improvements). Moreover, the entire chain begins with the vulnerability identification step, which undoubtedly employs specialized tools, given that manual evaluation is impracticable for large projects.

This section presents the key concepts and technologies related to LSVerifier, an automated approach for software project evaluation. We focus on its structure and implementation to analyze security vulnerabilities in open-source codebases.

\subsection{Bounded Model Checking Technique} 
\label{sec:BMC}

Bounded model checking (BMC) is a formal verification technique that detects errors up to a specified depth $k$, using Boolean Satisfiability (SAT) or Satisfiability Modulo Theories (SMT). Consequently, without a known upper bound for $k$, BMC cannot guarantee complete system correctness. In addition, as it only explores a limited state space by unwinding loops and recursive functions to a maximum depth, the state-explosion problem is inherently alleviated. In summary, this bounded nature makes BMC effective for uncovering fundamental errors in applications~\cite{Clarke, Gadelha2019}. Properties under verification are defined by
\begin{equation}
\text{{BM}}C_{\Phi}(k) = I(s_1) \land \left( \bigwedge_{i=1}^{k-1} T(s_i, s_{i+1}) \right) \land \left( \bigvee_{i=1}^{k} \neg \phi(s_i) \right),
\end{equation}

\noindent where $I(s_1)$ is the set of initial states for a system, $\bigwedge_{i=1}^{k-1} T(s_i, s_{i+1})$ is the transition relation between time steps $i$ and $i+1$, encompassing the evolution of the system over $k$ steps, and $\bigvee_{i=1}^{k} \neg \phi(s_i)$ represents the negation the property $\phi$ at state $s_i$, indicating its violation within a bound $k$. Together, these components formulate a problem that is satisfiable if and only if a counterexample of length $k$ or less exists, which includes the necessary information for its reproducibility. 

\subsection{LSVerifier Tool}
\label{sec:lsverifier}

The LSVerifier tool~\cite{de2023finding, de2023lsverifier} provides comprehensive support for the entire C$11$ standard, the current version of the C programming language. Moreover, unlike other tools based on SAST, such as Fortify SCA and Coverity, it can handle entire software projects and not only main entry functions, presenting high flexibility and coverage. It identifies software vulnerabilities by simulating a finite program execution prefix that includes all possible defined inputs, explicitly generating one symbolic execution per interleaving~\cite{cordeiro2011verifying}. 

\begin{figure*}[htb!]
 	\centering
 	\includegraphics[width=15cm]{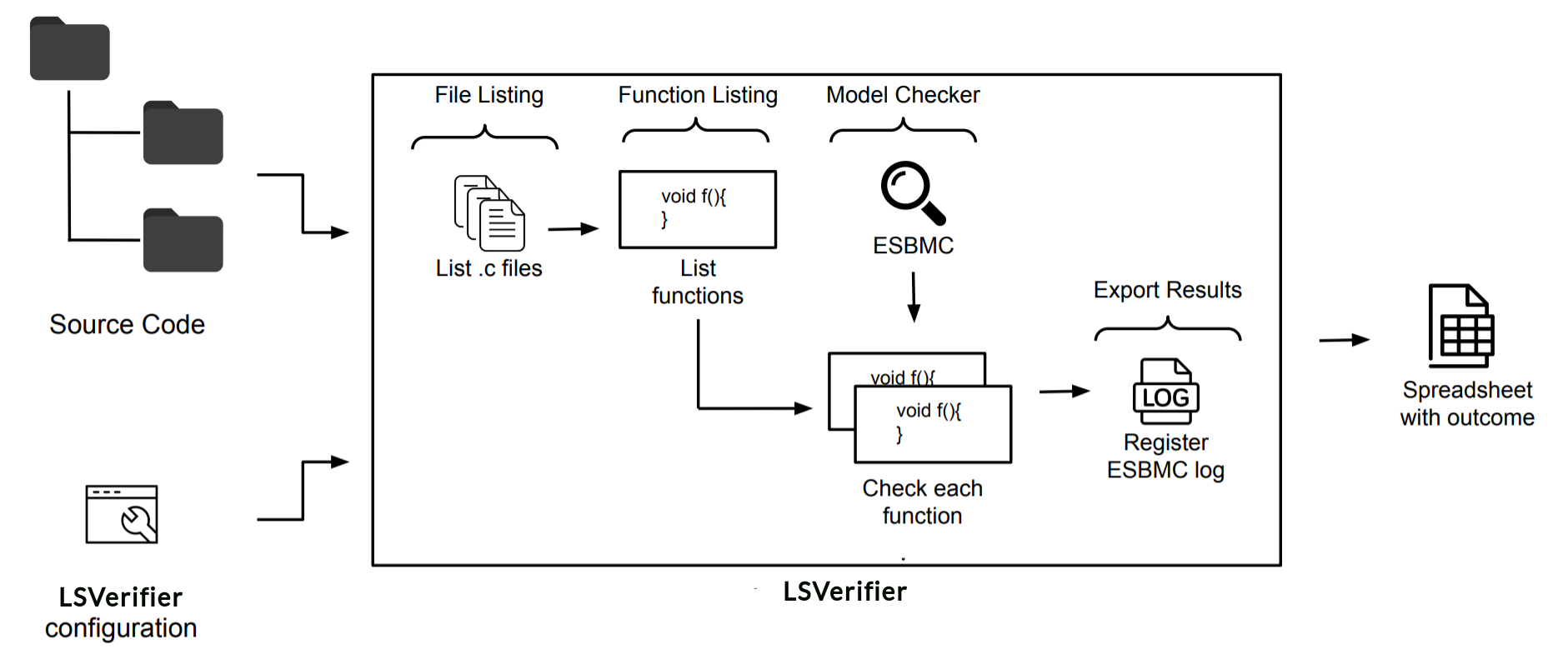}
 	\caption{The LSVerifier's verification process involves specifying source-code directory and configurations, including solver, encoding, and verification methods. Violations are categorized and reported in a detailed summary output.}
 	\label{script-new}
\end{figure*}

LSVerifier supports the detection of various vulnerabilities, including buffer overflows, arithmetic overflows, invalid pointer access, improper buffer access, null pointer dereferences, double frees, division by zero, array bounds violations, pointer arithmetic violations, and user-defined assertions. The verification process is illustrated in Figure~\ref{script-new}, which requires specifying the source code directory and configurations, such as the solver, encoding, and verification methods.

LSVerifier conducts a comprehensive verification process by specifying the target source-code directory and the required configuration, including solver, encoding, and verification methods. Subsequently, all \texttt{.c} files inside the input directory are listed and examined using the Efficient SMT-based Context-Bounded Model Checker (ESBMC)~\cite{gadelha2021esbmc}, leading to the creation of a report summarizing the obtained results. 

\begin{figure}[htb]
 	\centering
 	\includegraphics[width=15cm]{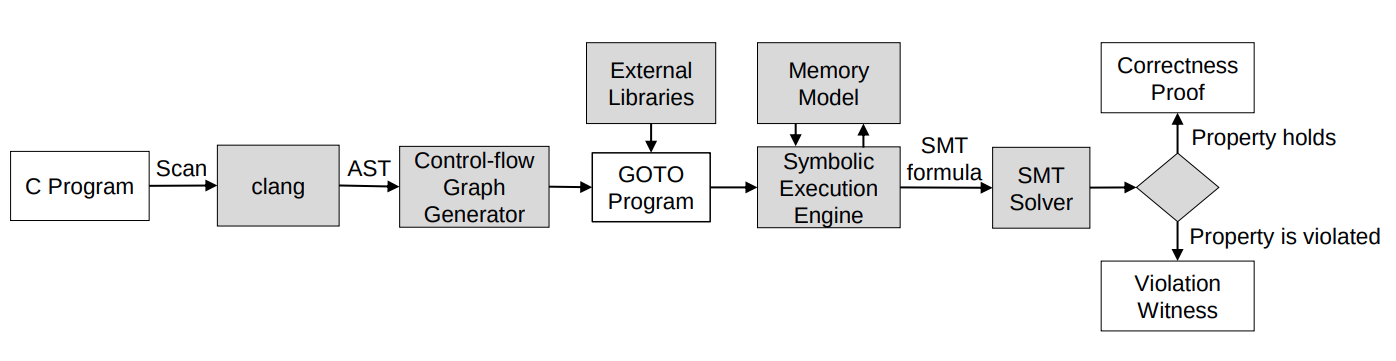}
        \caption{ESBMC verifier approach. White rectangles represent input and output and gray rectangles represent the verification steps~\cite{gadelha2021esbmc}.}
        \label{ESBMC_arch}
\end{figure}

The core BMC methodology employed by ESBMC involves unfolding a target system for a limited number of iterations and formulating a verification condition (VC). If the latter is satisfiable, it indicates a counterexample for a given property at a specific depth. ESBMC, in turn, is a robust and publicly available formal software verifier selected as our BMC module for formal verification. ESBMC employs state-of-the-art incremental BMC techniques and \textit{k}-induction proof-rule algorithms based on abstract interpretation, constraint programming (CP), and SMT solvers, whose effectiveness has already been demonstrated in various contexts~\cite{beyer2024state}. The ESBMC's architecture is illustrated in Figure \ref{ESBMC_arch}. Its core detection mechanism relies on BMC, which converts source code into formal logical representations. These formulae encode a program’s behavior and the associated properties to be verified, such as the memory-safety ones. The resulting encoded logic is then passed to an SMT solver, which systematically explores a program's state space.

ESBMC employs several key components during its verification process. The Control-flow Graph (CFG) Generator handles C++ programs by including type-checking and static analysis, creating an Intermediate Representation (IR) for GOTO program generation. At the same time, ANSI-C converts Abstract Syntax Trees (AST) into GOTO programs with additional checks and simplifications. The Symbolic Execution Engine symbolically executes the GOTO program, unrolling loops, generating Static Single Assignments (SSA) forms, and deriving safety properties for SMT solvers, including pointer safety checks. The SMT Back-end supports multiple solvers, encoding the SSA form into a formula to check satisfiability and generate counterexamples if a bug is detected.

Any property violations found during a verification procedure are informed and categorized by LSVerifier via a detailed report. For example, if a buffer overflow is detected, it flags the problematic function, highlighting the violated bounds, and generates a detailed report with the corresponding counterexample, which aids developers in understanding root causes. It includes a sequence of states and transitions, showing how a system evolves from an initial state to a condition where a property is violated. Indeed, this trace provides critical information for debugging as it pinpoints the exact sequence of operations leading to an error. Consequently, by analyzing a counterexample, developers can understand what causes a violation and then take corrective actions to fix its underlying issue.

\section{Methodology}
\label{sec:Methodology}

In this section, we present an overview of the verification methodology employed in this research, along with the experimental setup used to validate our approach.

\subsection{Vulnerability Detection Process}

\begin{figure}[htb!]
 	\centering
 	\includegraphics[width=14cm]{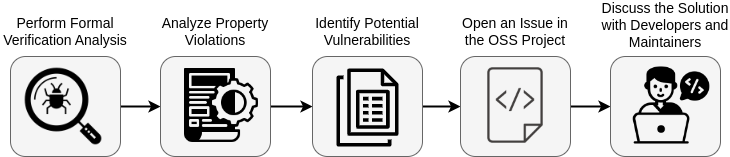}
        \caption{Verification methodology using LSVerifier.}
        \label{LSVerifier_Verification}
\end{figure}

In this section, we introduce the principles of our approach to detect the presence of vulnerabilities, based on the concepts previously introduced in our work. The verification procedure using LSVerifier is displayed in Figure \ref{LSVerifier_Verification}, where a formal verification process begins with a thorough analysis using specialized tools to ensure compliance with specified security properties. This way, violations are identified and categorized based on their nature and severity. Next, the associated potential vulnerabilities are assessed to confirm whether they represent real security threats. If a valid vulnerability is identified, an issue is opened in the respective OSS project's repository, providing detailed information about it. This process continues with discussions between the project's developers and maintainers to explore potential fixes and solutions for the identified issue, collaboratively.

The data collected from this verification methodology is used to address the research questions. To answer \textbf{RQ1}, this study begins by identifying and detailing the vulnerability types commonly found in OSS projects and also their prevalence, laying the foundation for understanding the nature of a property violation that leads to a possible software vulnerability. For \textbf{RQ2}, we explore how developers' actions, i.e., response to issues, maintenance of dependencies, and overall security practice, influence vulnerability mitigation, connecting human factors to security challenges. Finally, to address \textbf{RQ3}, we present the most effective strategies for mitigating risks from dependency vulnerabilities, offering actionable insights and solutions based on the identified problems and developers' behaviors.

\subsection{Experiment Setup}

All experiments described in this study were conducted on a system equipped with an Intel(R) Core(TM) i$7$-$9750$H computer processing unit (CPU) operating at $2.60$ GHz, using $32$ GB of RAM, and running Ubuntu $22.04$. For benchmarking purposes, we curated a dataset comprising ten widely used software modules written in C: VideoLAN Client (VLC) in version $3.0.18$, VI improved (VIM) in version $9.0.1672$, terminal multiplexer (Tmux) in version $3.3$a, reliable USB formatting utility (RUFUS) in version $4.1$, OpenBSD secure shell (OpenSSH) in version $9.3$, cross-platform make (CMake) in version $3.27.0$-rc$4$, network data (Netdata) in version $1.40.1$, Wireshark in version $4.0.6$, Open Secure Sockets Layer (OpenSSL) in version $3.1.1$, PuTTY in version $0.78$, structured query language lightweight (SQLite) in version $3.42.0$, and remote dictionary server (Redis) in version $7.0.11$. All open-source software utilized in this research was distributed under open-source licenses, including GNU GPL, Apache, and MIT.

The following command was used to run LSVerifier on the entire set of OSS projects to analyze the codebase and identify potential vulnerabilities:

\begin{footnotesize}
\begin{verbatim}
"$ lsverifier -r -f -l dep.txt".
\end{verbatim}
\end{footnotesize}
\noindent The parameter \texttt{-l dep.txt}  specifies a file containing paths for including header files from dependencies, ensuring that all necessary resources are considered. The parameter \texttt{-f} enables function verification, verifying individual functions within a codebase. Finally, the parameter \texttt{-r} enables recursive verification, ensuring that the verification process includes all nested functions and dependencies. 

\section{Empirical Study Results }
\label{cap:results}

This section presents the investigation results on vulnerabilities in OSS project dependencies and their impact on overall project security, highlighting how developers' behaviors and practices influence vulnerability mitigation.

\subsection{OSS Projects Exploitation}

The issues reported in this study were based on the counterexample traces provided by LSVerifier during its analysis procedures. In this context, OSS projects were assessed according to the methodology outlined in Section~\ref{sec:Methodology}. 

Therefore, Table~\ref{properties_analysis} provides an overview of the issues reported, analyzed, and fixed in the chosen OSS projects. It is worth noticing that such issues were discussed with the respective developers and maintainers of the chosen OSS projects, which enabled us to evaluate and confirm many of them.   

\begin{table*}[!htb]
  \centering
  \begin{minipage}{\textwidth}
    \centering
    \footnotesize
    \caption{Issues reported to the open-source software project repositories.}
    \begin{tabular}{|l|c|c|}
    \hline 
    \textbf{OSS project} & \textbf{Issues reported} & \textbf{Issues fixed} \\ 
    \hline
    VLC & Issue 1\footnote{\scriptsize\url{https://code.videolan.org/videolan/vlc/-/pipelines/227531}} & 1 \\
    \hline
    VIM & Issue 1\footnote{\scriptsize\url{https://github.com/vim/vim/issues/9571}} & 0 \\
    \hline
    RUFUS & Issue 1\footnote{\scriptsize\url{https://github.com/pbatard/rufus/issues/1856}}, Issue 2\footnote{\scriptsize\url{https://github.com/kokke/tiny-regex-c/issues/76}} & 1 \\
    \hline
    OpenSSH & Issue 1\footnote{\scriptsize\url{https://bugzilla.mindrot.org/show_bug.cgi?id=3452}}, Issue 2\footnote{\scriptsize\url{https://bugzilla.mindrot.org/show_bug.cgi?id=3382}} & 0 \\
    \hline
    CMake & Issue 1\footnote{\scriptsize\url{https://gitlab.kitware.com/cmake/cmake/-/issues/23132}} & 1 \\
    \hline
    Netdata & Issue 1\footnote{\scriptsize\url{https://github.com/netdata/netdata/issues/13219}}, Issue 2\footnote{\scriptsize\url{https://www.sqlite.org/forum/forumpost/3ffffb11d0}} & 0 \\
    \hline
    Wireshark & Issue 1\footnote{\scriptsize\url{https://gitlab.com/wireshark/wireshark/-/issues/17897}} & 1 \\
    \hline
    OpenSSL & Issue 1\footnote{\scriptsize\url{https://github.com/openssl/openssl/issues/17560}} & 0 \\
    \hline
    SQLite & Issue 1\footnote{\scriptsize\url{https://sqlite.org/forum/forumpost/ac645ab114}}, Issue 2\footnote{\scriptsize\url{https://www.sqlite.org/forum/forumpost/a2d232d413}} & 0 \\
    \hline
    Redis & Issue 1\footnote{\scriptsize\url{https://github.com/janislley/lsverifier_final_results/blob/main/redis-7.0.11/out/issue1.pdf}}, Issue 2\footnote{\scriptsize\url{https://github.com/janislley/lsverifier_final_results/blob/main/redis-7.0.11/out/issue2.pdf}} & 0 \\
    \hline
    \end{tabular}
    \label{properties_analysis}
  \end{minipage}
\end{table*}

\subsection{RQ1: What are the Common Types and Prevalence of Dependency Vulnerabilities in Open-Source Software Projects?}

Our evaluation of developer behavior in mitigating security vulnerabilities in OSS projects revealed crucial insights into current practices and their broader implications for fostering a trustworthy software ecosystem. 

In the VLC project, a pointer dereference issue was identified in the framebuffer third-party library.
A double-free error was identified as its cause, a vulnerability related to CWE-$415$~\cite{cwe}. Consequently, the respective maintainers decided to remove the Linux {\it fbdev} subsystem, which has been deprecated for over a decade, as superior alternatives are now available. This proactive approach led to an immediate impact on mitigating such vulnerabilities.

In our analysis of RUFUS, we identified property violations such as array bounds, division by zero, and invalid pointers.
Each issue highlights specific code errors and their implications, providing insights into the root causes and potential fixes for the identified vulnerabilities in RUFUS's software structure. However, when writing the present paper, we received only one bug fix for the library tiny-regex-c to address an out-of-bounds violation related to CWE-$787$~\cite{cwe}. Indeed, such behavior implies careless maintenance, a key aspect that can cause higher future impacts on system availability and reliability.

In the case of CMake,
developers promptly addressed a pointer dereferencing issue in the source code by adding a verification step before pointer usage, which was caused by an invalid pointer related to CWE-$824$~\cite{cwe}. This result highlights the importance of developers being aware of potential memory management issues and adopting defensive programming practices, such as boundary-checking on memory access operations. By prioritizing secure memory management practices, developers can mitigate serious security vulnerabilities in their projects.

In our investigation of Wireshark,
we uncovered common types of dependency vulnerabilities, including array access violations related to CWE-$125$ ~\cite{cwe}, and invalid and null pointers related to CWE-$824$ and CWE-$476$~\cite{cwe}, respectively. These vulnerabilities were identified in the CMake and network programming language (NPL) libraries, which are critical project dependencies. The issues stemmed from dereference failures caused by out-of-bounds access and null pointer occurrences. Notably, the library NPL has not been actively maintained as its last significant update occurred approximately nine years ago. Besides, the most recent commit log reference dates back eight years. This lack of maintenance highlights the prevalence and risk of dependency vulnerabilities in OSS projects. To mitigate such problems and ensure the robustness and security of Wireshark, the development team decided to remove the library NPL, which is a significant result.

\begin{tcolorbox}[title={Finding 1}, coltitle=white, colbacktitle=black]
Based on the vulnerabilities identified and mitigated in this study, common types of dependency vulnerabilities in open-source software projects include pointer dereference issues, such as the double-free errors (CWE-415) found in VLC; array access violations, including out-of-bounds violations (CWE-787) in RUFUS; invalid pointers detected in CMake and Wireshark (CWE-824); and null pointer dereferences identified in Wireshark (CWE-476). These findings demonstrate that such vulnerabilities are not isolated incidents but recurring issues in dependency management, confirming the need for more systematic and proactive mitigation strategies to ensure OSS project security.
\end{tcolorbox}
\label{finding1}

The widespread occurrence of these vulnerabilities highlights the significant security risks posed by faulty dependencies in OSS projects. It also emphasizes the importance of proactive and consistent management of third-party libraries to safeguard OSS security and stability.

\begin{tcolorbox}[title={Finding 2}, coltitle=white, colbacktitle=black]
Our findings highlight the recurring security challenges in OSS projects, particularly in managing third-party libraries. Developers' actions, such as removing deprecated subsystems and adding verification steps, demonstrate the critical role of proactive maintenance in mitigating security vulnerabilities. Indeed, such aspects underscore that continuous monitoring and management of dependencies are not merely best practices. They are also essential measures required to maintain the integrity and security of OSS projects.
\end{tcolorbox}
\label{finding2}

\subsection{RQ2: How do developers' behaviors and practices influence the mitigation of security vulnerabilities?}

Understanding how developers' behaviors and practices influence security vulnerability mitigation is crucial for enhancing OSS project security. Developer responses to the identified vulnerabilities, their approach to maintaining dependencies, and their willingness to adopt proactive security measures play a significant role in mitigating risks. 

Diligent maintenance is crucial in OSS projects, as it ensures the timely addressing vulnerabilities, bug fixes, and compatibility updates. Neglecting these responsibilities can lead to unaddressed security flaws, reduced system performance, and increased risk of system failures. Consistent maintenance practices are essential to sustain OSS projects' security and reliability, ensuring they remain robust and dependable for users.

The SQLite project's response to identified violations underscores a prevalent challenge in the software development community: the inclination to dismiss static analyzer or formal verifiers results as ``false positives''.
This practice stems from the belief that static analyzers often produce inaccurate results, causing unnecessary alarms and potentially wasting development resources. The SQLite team highlighted that they usually disregard these reports without concrete evidence, such as an SQL script or specific code reproducing the issue. While pragmatic and aimed at preventing undue alarm, this stance carries significant risk. By dismissing these warnings, the team may overlook potential vulnerabilities that have not yet manifested and could be avoided. Indeed, it reveals the usual paradigm: corrections only arrive after a real problem, which shows a lack of proactivity and leads to higher losses.

The reliance on historical codebase performance further exacerbates this issue. During our discussions, the SQLite team noted their confidence in their codebase's historical stability, which they believe confuses static analyzers. This over-reliance can lead to complacency, resulting in missed opportunities to address latent issues before they become significant security threats. By not investigating potential false positives, in their opinion, developers may inadvertently leave their software susceptible to vulnerabilities that are initially difficult to detect but could have severe implications if exploited.

Besides, such an approach highlights a critical gap in development processes: the need for a balanced view of static analysis results. On the one hand, while it is true that not all warnings require immediate action, completely disregarding them without thorough investigation can undermine the overall security of a given software system. Thus, a more nuanced approach, where static analyzers' findings are carefully evaluated and verified, can help identify genuine issues early, thus enhancing software security and robustness. 

\begin{tcolorbox}[title={Finding 3}, coltitle=white, colbacktitle=black]
Our analysis reveals a gap in the development process related to the interpretation of static analysis results. Although static analyzers may generate false positives, they often identify legitimate issues that may be missed during manual code reviews. In contrast, formal verifiers, supported by mathematical proofs, ensure higher accuracy. Thus, developers must integrate both tools, balancing skepticism and due diligence, to enhance software systems' overall security and reliability.
\end{tcolorbox}
\label{finding3}

Similarly, the developers acknowledged an issue reported in OpenSSL involving an invalid pointer dereference related to CWE-$476$~\cite{cwe}, but not classified as a vulnerability. It happened because many OpenSSL APIs crash if a null pointer is passed
However, this perspective reveals a problematic practice: developers frequently assume that certain conditions will never occur, dismissing potential vulnerabilities, which can be dangerous. If an attacker manipulates parameters or code to create these conditions, even using regular code contribution tools, the identified problem could lead to severe consequences, including system crashes or security breaches.

This example highlights the importance of changing the usual behavior of developers to make them address all identified issues, regardless of associated perceived likelihood. By not considering these scenarios as potential vulnerabilities, it is clear that developers leave their code open to exploitation. Addressing seemingly unlikely issues can prevent attackers from leveraging them to compromise the system. Besides, encouraging a proactive approach to vulnerability management, where all identified issues are investigated and resolved, is essential for maintaining robust security.

Moreover, this practice underscores the need for developers to anticipate and mitigate even rare scenarios. This shift in behavior involves recognizing that assumptions about the improbability of certain conditions can lead to significant security gaps. A comprehensive approach to security should include evaluating and addressing all potential issues and ensuring that a software structure is resilient against a wide range of attacks. In conclusion, fostering a culture of thorough investigation and resolution of all identified vulnerabilities is crucial for the security and integrity of OSS projects.

\begin{tcolorbox}[title={Finding 4}, coltitle=white, colbacktitle=black]
Our findings show that dismissing potential issues, such as buffer overflows and dereference failures identified by static analysis or formal verification (e.g., model-checking) tools, without proper investigation, leaves software vulnerable to real threats. Such results emphasize the importance of adopting a balanced approach that integrates both manual testing and static analysis to ensure robust security in open-source C projects.
\end{tcolorbox}
\label{finding4}

\subsection{RQ3: What is the most effective strategy for mitigating risks from dependency vulnerabilities in open-source software projects?}

As our analysis indicates, the most effective strategy for mitigating risks from dependency vulnerabilities, in OSS projects, is to reduce the number of direct dependencies. It can be achieved by carefully selecting and substituting multiple small libraries with a single and well-established element known for its strong security track record. This approach simplifies dependency management and leverages widely used and reputable open-source libraries' security practices and community support. Although it does not dismiss thorough analysis and careful evaluation, as previously suggested, it may reduce risks and also the revision workload.

The analysis of Redis revealed multiple violations,
including array bound violations, related to CWE-$787$,
invalid pointer dereferences, related to CWE-$476$~\cite{cwe}, null pointer dereferences, related to CWE-$476$~\cite{cwe}, and out-of-bounds object access, related to CWE-$119$~\cite{cwe}. While some of these were confirmed as false positives, a significant oversight was identified: inadequate null pointer checks. Indeed, this oversight could lead to undefined behavior if a function is called with a null pointer. Even so, the Redis developers dismissed this issue, claiming that the function or method would never be invoked in a problematic way. This is a dangerous assumption, as attackers could potentially exploit such scenarios. In addition, it is worth noticing that these issues were not false positives but problems dismissed by wrong assumptions made during the development or verification process.

\begin{tcolorbox}[title={Finding 5}, coltitle=white, colbacktitle=black]
This finding highlights the critical need for thorough verification of false positives, as dismissing them without adequate investigation can lead to overlooked vulnerabilities that impact the security of the software. Ensuring that potential false positives are rigorously checked and validated is essential to prevent security weaknesses from being inadvertently introduced into the codebase.
\end{tcolorbox}
\label{finding5}

These oversights could have critical consequences, particularly in C programs, which often lack robust memory management and are more susceptible to vulnerabilities such as buffer overflows, null pointer dereferences, and memory leaks. Failing to identify and address these overlooked vulnerabilities leaves the system exposed to exploitation, where attackers can manipulate these memory issues to compromise security. Strengthening overall system security requires a meticulous approach to detecting and resolving these potential weaknesses in memory handling.

\begin{tcolorbox}[title={Finding 6}, coltitle=white, colbacktitle=black]
Our analysis indicates that functions from dependency libraries, especially in C programs, where pointers are frequently used to access arrays, pose serious security risks if not carefully verified. It highlights the inherent vulnerability in passing pointers as function arguments, which can lead to significant security concerns when not properly addressed during development processes.
\end{tcolorbox}
\label{finding6}

Results indicate that managing dependency vulnerabilities in OSS projects is more effective when reducing direct dependencies rather than expanding development teams. This can be achieved in OSS projects by carefully selecting and replacing multiple smaller libraries with a single and well-established library known for its robust security track record. Such an approach simplifies dependency management and leverages widely used and reputable open-source libraries' security practices and community support. 

\begin{tcolorbox}[title={Finding 7}, coltitle=white, colbacktitle=black]
Our results demonstrate that effective library management plays a more crucial role in mitigating dependency vulnerabilities in OSS projects than increasing the number of contributors, project activity, or overall project size. The associated analysis reveals that reducing the number of direct dependencies, such as replacing several smaller libraries with a single and well-established element with a strong security record, is a critical factor in enhancing software security.
\end{tcolorbox}
\label{finding7}

By prioritizing the integration of libraries that have a proven track record of security, reliability, and consistent updates, developers can significantly reduce the risk of introducing vulnerabilities into their software. These well-vetted libraries typically undergo extensive peer review and real-world testing, making them more resilient to attacks. 

Furthermore, selecting libraries with active maintenance ensures that any newly discovered security flaws are promptly addressed through patches and updates, minimizing exposure to potential threats. Incorporating such trusted libraries into the development process also allows developers to focus more on their core application logic rather than spending excessive time identifying and fixing third-party code vulnerabilities. This approach not only mitigates common security risks but also ensures that the software remains resilient against emerging threats in an ever-evolving threat landscape.

\section{Conclusion} 
\label{cap:conclusion}

Our findings emphasize the need for developers to adopt a more rigorous approach to security, particularly regarding third-party libraries. Despite their potential for false positives, static analyzers (SAST an SCA) play a crucial role in identifying genuine issues that may be missed during manual reviews. The formal verification approach implemented by LSVerifier provides detailed reports with counterexamples, which can help developers ensure code safety and improve the security and resilience for their software systems. Developers must also balance addressing these reports with a collaborative approach, working with security researchers and tool developers to validate and fix potential vulnerabilities. By fostering a culture that prioritizes security and encourages thorough examination of all potential risks, the open-source community can enhance software projects' overall integrity and robustness.

This study demonstrated the effectiveness of the proposed mitigation strategies, leading to the successful resolution of four vulnerabilities in the OSS projects VLC, RUFUS, CMake, and Wireshark. Such results underscore the critical role of proactive dependency management in enhancing software security. By addressing our three research questions, we have identified key best practices that developers and the OSS community can adopt to strengthen security measures significantly, as follows:

\begin{itemize}
    \item providing comprehensive dependency management;
    \item integrating formal verification tools;
    \item fostering a security-first culture;
    \item using well-established libraries;
    \item enforcing regular security audits and reviews.
\end{itemize}

This study emphasizes the importance of actively managing project dependencies to prevent security risks. Indeed, developers should focus on minimizing the number of direct dependencies and thoroughly auditing both direct and transitive dependencies. In addition, our results highlight the need for integrating static analysis and formal verification tools into development processes. By using tools based on formal verification that can detect deeper issues, such as memory management flaws, comprehensive security can be assured, complementing traditional static analysis methods. Also, in this context, Maintaining robust security in OSS projects requires a culture that prioritizes security throughout the entire development lifecycle. 

Therefore, developers must collaborate with security experts to efficiently detect and resolve vulnerabilities, and the OSS community can better protect against emerging threats by encouraging security-first practices. Moreover, developers and maintainers can significantly mitigate risks by focusing on well-established libraries with strong security records. So, reducing reliance on poorly maintained or obscure libraries helps minimize vulnerabilities and improve project security, as demonstrated by the successful resolution of issues in major OSS projects. Finally, Regular security audits and rigorous code reviews are essential for maintaining a strong security posture in OSS projects. Consequently, this study reinforces the importance of adopting a zero-trust culture to ensure that all issues are thoroughly analyzed and addressed.

Maintaining robust security in OSS projects requires a multifaceted approach that combines static analysis, formal verification, and collaborative efforts with security experts. Developers can significantly lower security risks by reducing unnecessary dependencies, selecting well-vetted libraries, and continuously monitoring and managing dependencies. The OSS community must prioritize security by adopting best practices, enforcing regular updates, and remaining vigilant against emerging threats, given that analysis emphasizes the critical importance of diligent maintenance.

Besides, OSS projects can evolve into more resilient and trustworthy software ecosystems by adopting integrated approaches where security, testing, evaluation, and analysis are regarded with the same importance as development activities. Neglecting regular updates and vulnerability management can lead to severe consequences. Therefore, incorporating agile strategies with thorough testing, evaluation, and analysis throughout development lifecycles will improve the robustness and dependability of OSS projects. In summary, fostering a security-conscious mindset and embedding best practices into the development process is essential for ensuring the security and longevity of OSS projects.

\section*{Acknowledgment} 
\label{cap:agradecimentos}

The authors are grateful for the support offered by the SIDIA R\&D Institute in the SEICO project. Samsung partially supported this work, using Informatics Law resources for Western Amazon (Federal Law No. 8.387/1991). Therefore, the present work disclosure is in accordance as foreseen in article No. 39 of number decree 10.521/2020.
The work in this paper is also partially funded by the Engineering and Physical Sciences Research Council (EPSRC) grants EP/T026995/1, EP/V000497/1, EP/X037290/1, and Soteria project awarded by the UK Research and Innovation for the Digital Security by Design (DSbD) Programme.

\bibliographystyle{sbc}
\bibliography{sbc-template}

\end{document}